# Simulation for field emission images of micrometer-long SWCNTs

**Weiliang Wang, Ningsheng Xu, Zhibing Li**[*]

State Key Laboratory of Optelectronic Materials and Technologies and

School of Physics and Engineering, Sun Yat-Sen University, Guangzhou 510275, China

**Abstract:**

*The electron distribution of open-ended single-walled carbon nanotubes with chirality indexes (7,0) and (5,5) in the field emission conditions was calculated via a multi-scaled algorithm. The field emission images were produced numerically. It was found that the emission patterns change with the applied macroscopic field. Especially, the symmetry of the emission pattern of the (7,0) carbon nanotube is breaking in the lower field but the breaking is less obvious in the higher field. The enlargement factor increases with the applied macroscopic field.*

**Keyword**: *carbon nanotube, field emission, image*

---

[*] Corresponding author: stslzb@mail.sysu.edu.cn

# Ⅰ. INTRODUCTION

The carbon nanotubes (CNTs) can emit electrons at low applied macroscopic fields (AMFs). This phenomenon has attracted many experimental and theoretical studies. The process is basically quantum tunnelling of electron from the CNT to vacuum through a potential barrier at the apex. Therefore it is very sensitive to the AMF and the electronic structure of the apex. Field emission (FE) microscopy had been used to investigate the structure of CNTs in atomic resolution.[1] The FE patterns were found to change with the AMF.[2] Many efforts had been dedicated to the calculation of the FE patterns.[3-5] Particularly, Ref. [3] has shown that it is possible to determine the configuration of the caps through experimental FE patterns. To take full advantage of FE pattern, it is essential to acquire a quantitative connection between the observed FE image and electronic structure of the tip. Our study aimed to show the possibility of quantitative simulation for the process and to demonstrate that the FE images do contain detailed information of the electronic structure.

In principle, the electronic structure depends on the atomic geometry of the whole CNT and also on the strength of the AMF. It should be obtained by quantum mechanical methods that are capable to cover both atomic scale (angstrom) and the scale of the CNT length (micrometer). On the other hand, the image is detected on a screen that locates in a macroscopic distance from the CNT.[1,2,6] The FE pattern will undergo deformation in the process and form on the screen the enlarged image that depends on the electron energy potential in the whole region, which eventually is related to the electronic structure and the AMF. Therefore, a multi-scale algorithm is appealed in order to simulate the process.

# Ⅱ. CALCULATION METHOD

In the present paper, we should simulate the single-walled CNT (SWCNT) of 1.μm long under the AMFs to obtain the electron density distribution of each energy state near the Fermi level via a molecular/quantum mechanical method. [7,8] The effective work function lowering due to the field penetration [1] had been taken into account. With the electron density distribution on hand, the electron energy potential in the concerned region was obtained. The space region of the potential barrier is bounded by the inner turning surfaces and the outer turning surface where the kinetic energy of motion along the field direction is zero. Fig. 1a schematically illustrates the inner and outer turning surfaces of the potential barrier and the FE image on the screen (the scale is not realistic). The electron potential energy in the region near the tip of the open-ended (5,5) SWCNT, in the AMF 8.V/μm, is presented in Fig.1b. To keep the image clear, the core potential is cut at −5.08 eV. The dashed curve is the outer turning surface of electron in the Fermi level. The inner turning surfaces locate at the edges of the dark blue region of atomic cores.

In order to produce the FE image projected on a screen that is set in 1. cm away from the SWCNT apex, we introduced three nesting grids: (1) a fine grid covering the inner turning surfaces, with grid spacing 0.01 nm; (2) a middle grid covering the potential barrier, with grid spacing 0.1 nm; (3) a large grid covering the region in front of the apex where the field is not uniform, with grid spacing 1. nm. In Fig.1b, The dotted box is the intersection of the fine grid region. The middle grid region is a box of length 18. nm that is partly shown in Fig.1b as the region above the horizontal black line. The large grid region is a box with length 50 nm, also above the horizontal black line. The inner turning surface is discretized by the fine grid into small surface elements. The FE current from the i-th surface element of the inner turning surface was calculated by the method developed by Khazaei et al.,[4]

$$j_i(\omega) = \frac{2e\hbar}{m_e} f(\omega) S_i \lambda_i^{-2}(\omega) D_i^2(\omega) g_i(\omega, x'_{l,i})$$

Where $m_e$ is the electron mass; $f(\omega)$ is the Fermi-Dirac distribution for the energy $\hbar\omega$; $D_i^2(\omega)$ is the tunnelling probability along the most probable emission direction (MPED) [8] from the i-th surface element (the coordinate along this direction is denoted by $x'$); $g_i(\omega, x'_{l,i})$ is the local density of states (LDOS) at the i-th inner turning point $x'_{l,i}$; $\lambda_i(\omega)$ is given by

$$\lambda_i(\omega) = (\pi/3)^{1/2} (c_i/3)^{-1/3} \left[\Gamma(2/3)\cos(\pi/6)\right]^{-1}$$

where $c_i$ is obtained by fitting $(2m_e/\hbar^2)(u(x')-\omega)$ to $c_i^2(x'-x'_{l,i})$ at the inner turning point, with $u$ the potential. To find the MPED, the transmission probabilities along various directions were calculated in the JWKB approximation, within the middle grid. As a good approximation, one may assume that the electrons come out of the potential barrier move under the electric field classically. The calculation for the trajectories through the non-uniform field was accelerated by use of the large grid.

Ⅲ. RESULTS

In principle, all energy states should be taken into account. But as a result of the calculation, only the FE of the states near the Fermi level is important. According to the qusi-equilibrium approximation, the Fermi level of the SWCNTs is the same as that of the substrate, which is set to be -5.08 eV. Fig. 2 shows the isosurfaces of LDOS of states near the Fermi level of SWCNTs under the AMF of 8. V/μm and 18. V/μm. Remarkably, the axial rotation symmetry of the (7,0) SWCNT is breaking in the electronic structure (the top two rows of Fig.2). We suggested that is the consequence of the Jahn-Teller effect originated from the coupling of the electron edge states and the atomic structure distortion. In the third and fourth rows of Fig.2, the LDOS of the (5,5) SWCNT that has no edge state exhibits clear axial symmetry.

The LDOS changes with the AMF, [9] so does the FE pattern on the screen (Fig. 3). The symmetry breaking of (7,0) SWCNT in 8. V/μm is observable in the screen image. The symmetry breaking is less obvious in 18. V/μm, which can be explained via the duality of LDOS of two degenerate states (the left and the middle panels of the second row of Fig.2). In 18. V/μm, the main contribution comes from that pair of degenerate states and the average electron density of them possesses the symmetry. The FE images on the screen are larger under higher AMF. The enlargement factor is estimated to be $10^6$ for 8. V/μm and $10^7$ for 18. V/μm. The reason is that the emission direction is more confused to the thinner potential barrier in the higher field. The distribution of emission current from the inner turning surface elements in the emission angle against the tube axis is plotted in Fig.4 for the (5,5) SWCNT in 8. V/μm (black boxes) and 18. V/μm (circles) respectively. The emission angles are obviously larger in average in the higher field.

## Ⅳ. SUMMARY

In summary, the FE patterns of SWCNTs have been simulated via the multi-scale algorithm. The FE images reflect both electronic and atomic structures. The enlargement factor of $10^6$ ($10^7$) is found for the AMF 8. V/μm (18. V/μm) on the screen 1.cm apart from the apex of SWCNTs. The axial rotation symmetry breaking of the (7,0) SWCNT should be observable. The FE pattern changes with AMFs, revealing the change of symmetry properties of different electronic states in the tip.

## ACKNOWLEDGMENTS

Acknowledgement: the project is supported by the National Natural Science Foundation of China (Grant Nos. 10674182, 90103028, and 90306016) and National Basic Research Program of China (2007CB935500).

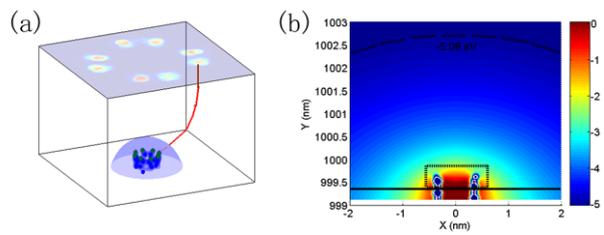

*Fig. 1.(color online) (a) Schematic illustration of the inner and outer turning surfaces of the potential barrier and the FE image on the screen (the scale is not realistic). The surfaces of green and blue balls are the inner turning surfaces. The larger light blue surface is the outer turning surface. The red line indicates a typical trajectory of an emitted electron. (b) Potential plot of the open-ended SWCNT in 8. V/μm. To keep the image clear, the core potential is cut at -5.08 eV. The dotted box is the intersection of the fine grid; the middle grid and the large grid cover the region above the horizontal black line. The dashed curve indicates the outer turning surface of electron in the Fermi level.*

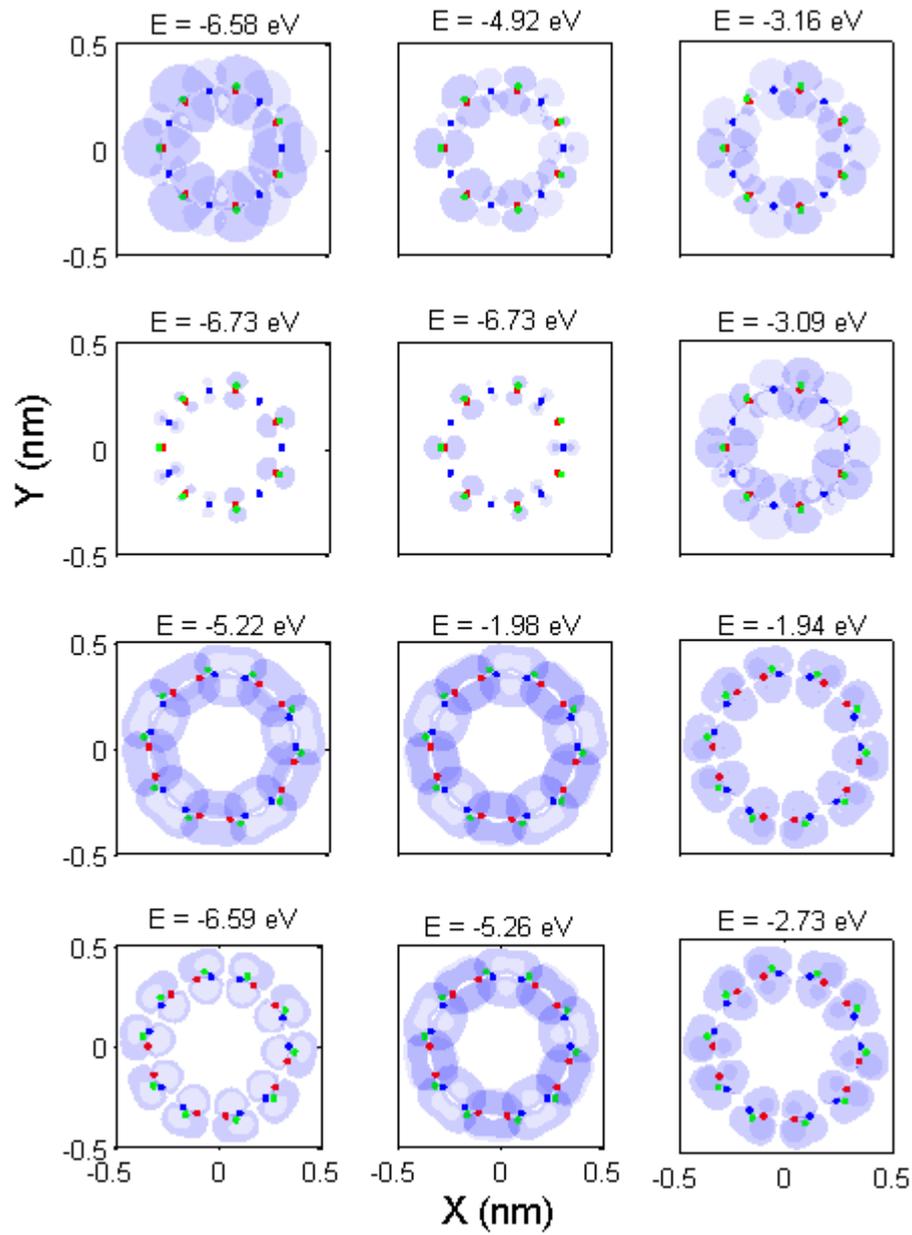

Fig. 2.(color online) From top to bottom: the first (second) row presents the isosurface of local density of state (LDOS) of three states near the Fermi level of (7, 0) SWCNT under 8. V/μm (18. V/μm). The third (fourth) row presents the LDOS of (5, 5) SWCNT under 8. V/μm (18. V/μm). The green circles indicate positions of H atoms; the red circles indicate positions of the first layer C atoms; the blue circles indicate positions of the second layer C atoms.

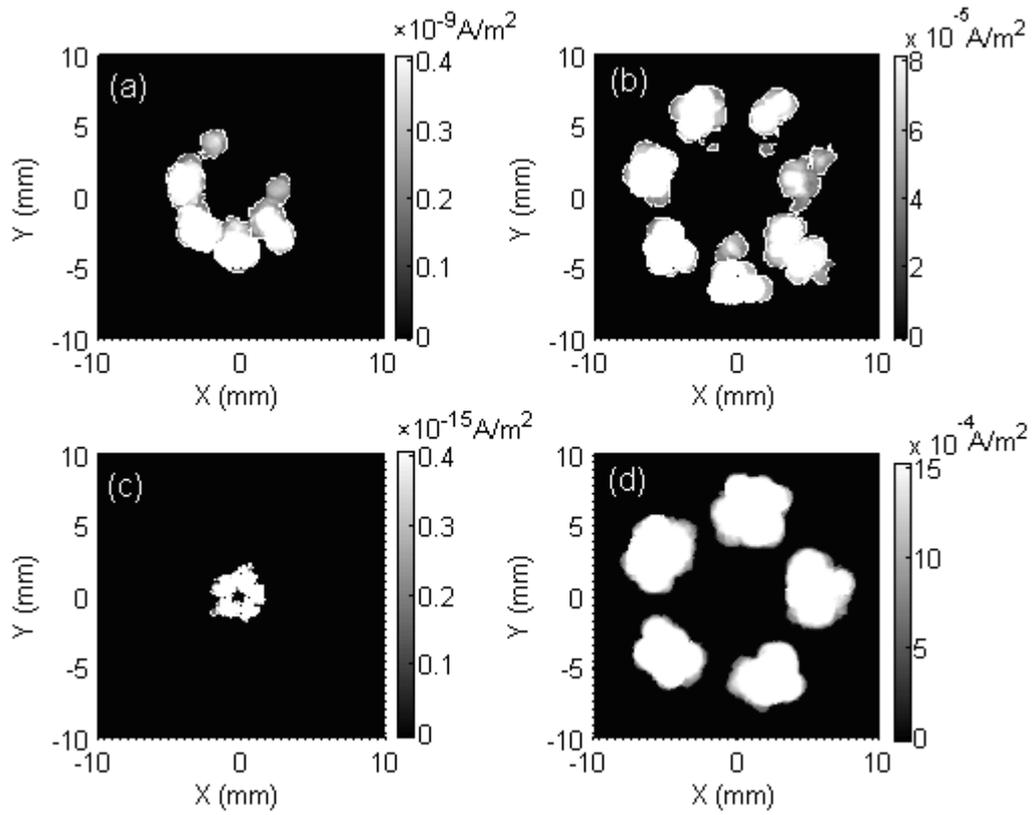

*Fig. 3.(color online) FE images on the screen for (7, 0) SWCNT under 8. V/μm (a) and 18. V/μm (b); FE images on the screen for (5, 5) SWCNT under 8. V/μm (c) and 18. V/μm (d).*

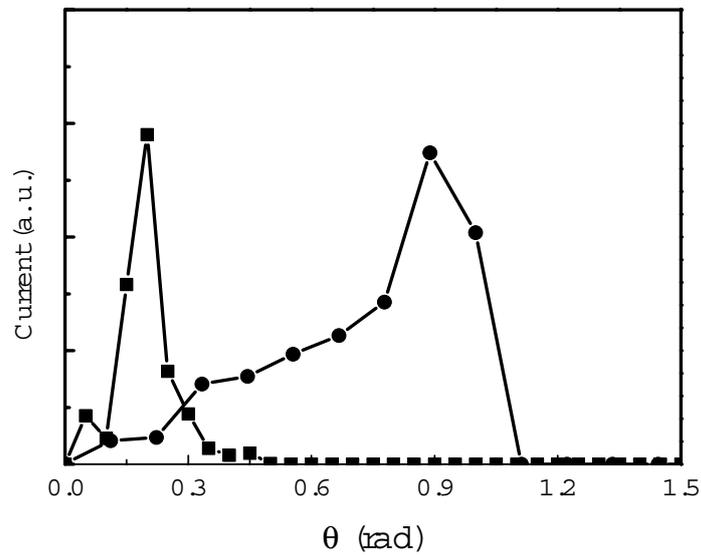

*Fig.4. The current distribution against emission angle from the inner surface elements. Black boxes for 8. V/μm and circles for 18. V/μm.*